%% file: ICHEP2018_NMSSMInflation.tex
\documentclass{PoS}

\include{paperdef}

\hypersetup{hidelinks}

\title{Phenomenological consequences of Higgs inflation in the NMSSM at the electroweak scale}

\ShortTitle{Phenomenology of Higgs inflation in the NMSSM at the electroweak scale}

\author{Wolfgang Gregor Hollik${}^{\,\dagger\,a,b,c}$, Stefan Liebler${}^{\,\ddagger\,d}$, Gudrid Moortgat-Pick${}^{\,\diamond\,a,e}$, \speaker{\mbox{Sebastian Pa{\ss}ehr}}$^{\,\, f}$, Georg Weiglein${}^{\,\circ\,a}$\\\ \\
${}^a$Deutsches Elektronensynchrotron, Notkestra{\ss}e 85, D--22607 Hamburg, Germany\\
\mbox{${}^b$Institute for Nuclear Physics, Karlsruhe Institute of Technology, D--76012 Karlsruhe, Germany}\\
${}^c$Institute for Theoretical Particle Physics, Karlsruhe Institute of Technology,\\
$\phantom{{}^c}$D--76128 Karlsruhe, Germany\\
\mbox{${}^d$Institute for Theoretical Physics, Karlsruhe Institute of Technology, D--76131 Karlsruhe, Germany}\\
${}^e$II. Institut f\"ur Theoretische Physik, Universit\"at Hamburg,\\
$\phantom{{}^e}$Luruper Chaussee 149, D--22761 Hamburg, Germany\\
$^f$Sorbonne Université, CNRS, Laboratoire de Physique Théorique et Hautes Énergies,\\
$\phantom{{}^f}$4 Place Jussieu, F--75252 Paris CEDEX~05, France\\\ \\
${}^\dagger$\email{w.hollik@desy.de}\\
${}^\ddagger$\email{stefan.liebler@kit.edu}\\
${}^\diamond$\email{gudrid.moortgat-pick@desy.de}\\
${}^\ast$\email{passehr@lpthe.jussieu.fr}\\
${}^\circ$\email{georg.weiglein@desy.de}
}

\abstract{
  
The Next-to-Minimal Supersymmetric Standard Model~(NMSSM) can
incorporate inflation, where a combination of the Higgs-doublet fields
plays the role of the inflaton. At the high scale, the Higgs doublets
are non-minimally coupled to supergravity; this coupling appears as an
additional contribution to the~$\mu$~term in the low-energy effective
superpotential and potentially changes physics at the electroweak
scale.

In a recent publication, we investigate the extended parameter space
of this model with respect to collider phenomenology at the
electroweak scale, and discuss scenarios which are potentially
different from the pure~NMSSM. We analyse the stability of the
electroweak vacuum, the masses of neutralinos/charginos and Higgs
bosons as well as the mixing and decays of Higgs bosons. Some
important aspects of this study are described in the following.

}

\FullConference{The 39th International Conference on High Energy Physics (ICHEP2018)\\
		4-11 July, 2018\\
		Seoul, Korea}

\hypersetup{colorlinks=true}

\begin{document}

\paragraph{Introduction}

During the period of inflation the universe has exponentially
increased its size. The idea of using the Higgs field of the Standard
Model~(\sm{}) as inflaton has been proposed
in~\citeres{Bezrukov:2007ep,Bezrukov:2008ej,Bezrukov:2009db}. However,
the simplest implementation is fine-tuned~\cite{Barbon:2009ya}. A
possible solution is given by the scale-free extension of the~\sm{} in
canonical superconformal supergravity models as proposed
by~\citeres{Ferrara:2010yw,Ferrara:2010in} based on earlier work
by~\citere{Einhorn:2009bh}. In these models, two Higgs
$SU(2)$-doublet~superfields~$\hat{H}_{u,d}$ are non-minimally coupled
to Einstein gravity via~$\chi\,\hat{H}_u\cdot\hat{H}_d$ with a
dimensionless constant~$\chi$; an additional Higgs singlet
superfield~$\hat{S}$ stabilises the potential during inflation,
see~\citere{Einhorn:2009bh}. These requirements can be implemented in
the $\mathbb{Z}_3$-invariant~\nmssm{} augmented by an additional
$\mu$~term, which we call~\munmssm{} in the following. In
\citere{Hollik:2018yek}, we study the low-energy electroweak
phenomenology of this model; studies at the scale of inflation are
given in \citeres{Ferrara:2010yw,Ferrara:2010in,Lee:2010hj}. We have
generated a model file for \FA~\cite{Kublbeck:1990xc, Hahn:2000kx},
\FC~\cite{Hahn:1998yk} and \LT~\cite{Hahn:1998yk}, where
\code{SARAH}~\cite{Staub:2009bi, Staub:2010jh, Staub:2012pb,
  Staub:2013tta} has been used to generate the tree-level couplings of
the~\munmssm{}, and at the one-loop order we have implemented the
counterterms and a renormalisation scheme that is compatible with the
schemes of~\citeres{Fritzsche:2013fta,Domingo:2017rhb} for the~\mssm{}
and~\nmssm. Leading two-loop corrections from the~\mssm{} are added
with the help
of~\FH~\cite{Heinemeyer:1998np,Heinemeyer:1998yj,Degrassi:2002fi,Frank:2006yh,Hahn:2010te,Bahl:2016brp,Bahl:2017aev,Bahl:2018FH}. The
compatibility of our scenarios with the experimental data is evaluated
with the help of~\HBv{5.1.0beta}~\cite{Bechtle:2008jh, Bechtle:2011sb,
  Bechtle:2013gu, Bechtle:2013wla, Bechtle:2015pma}
and~\HSv{2.1.0beta}~\cite{Bechtle:2014ewa}. In addition, we check the
stability of the electroweak vacuum with respect to non-standard
global minima. In view of the hints in the existing
data~\cite{Schael:2006cr,CMS:2017yta,ATLAS-CONF-2018-025}, we
investigate scenarios with light Higgs singlets with production
cross-sections taken from the~\nmssm{} version
of~\texttt{SusHi}~\cite{Harlander:2002wh,Harlander:2003ai}.

\paragraph{Analysis}

The superpotential of the~\munmssm{} with the non-minimal coupling to
supergravity reads $\mathcal{W} = \lambda\,\hat{S}\,\hat{H}_u\cdot
\hat{H}_d + \frac{1}{3}\,\kappa\,\hat{S}^3 +
\frac{3}{2}\,m_{3/2}\,\chi\,\hat{H}_u\cdot\hat{H}_d$ with the
gravitino mass~$m_{3/2}$. When the singlet field acquires a vacuum
expectation value~(vev)~$v_s$, the effective
parameter~\mbox{$\mue=\lambda\,v_s$} is generated. The value of~$\chi$
can be set to about~$10^5\,\lambda$~\cite{Lee:2010hj,Ferrara:2010in},
thus that the value of~\mbox{$\mu=\frac{3}{2}\,m_{3/2}\,\chi$} is
mainly steered by the gravitino mass: values of~\mbox{$\mu\gtrsim
  1$\,TeV} and~\mbox{$\lambda\gtrsim 0.1$} require light gravitinos
of~\mbox{$m_{3/2}\sim 100$\,MeV} (typically LSP), while
for~\mbox{$\lambda\lesssim 10^{-5}$} the gravitino mass is similar
to~$\mu$. The potential cosmological gravitino
problem~\cite{Moroi:1993mb}, where the light gravitino dark matter
overcloses the universe~\cite{Pagels:1981ke, Weinberg:1982zq}, can be
avoided by a low reheating
temperature~\cite{Ellis:1984eq,Khlopov:1984pf}. The Lagrangian
\mbox{$-\mathcal{L}_{\text{s}} = \left[A_\lambda\,\lambda\,S\,H_u\cdot
    H_d + \frac{1}{3}\,A_\kappa\,\kappa\,S^3 + B_\mu\,\mu\,H_u\cdot
    H_d + \text{h.\,c.}\right] + m_{H_d}^2\,\lvert H_d\rvert^2 +
  m_{H_u}^2\,\lvert H_u\rvert^2 + m_s^2\,\lvert S\rvert^2$}, which
softly breaks SUSY and the $\mathbb{Z}_3$-symmetry, is used. Further
parameters which break $\mathbb{Z}_3$-symmetry are generated
radiatively and renormalised in the \drbar{} scheme at the scale of
the top-mass~$m_t$. An extensive analytic discussion of the Higgs
potential and masses, vacuum stability, trilinear Higgs couplings, and
neutralino, chargino and sfermion masses is given
in~\citere{Hollik:2018yek}. The most important aspects are summarised
in the following: (1)~the Higgs mass matrix at the tree level contains
terms~${\propto}(\mu+\mue)$, $\kappa\,\mue$ and~$\mue^{-1}$ in the
singlet--doublet mixing, and terms~${\propto}\kappa\,\mue$,
$(\kappa\,\mue)^2$, $\mue^{-2}$ and~$\mu/\mue$ in the singlet
elements; (2)~full one-loop and leading MSSM two-loop corrections
of~$\mathcal{O}{\left(\alpha_t\alpha_s\right)}$~\cite{Heinemeyer:2007aq}
and~$\mathcal{O}{\left(\alpha_{t}^2\right)}$~\cite{Hollik:2014wea,Hollik:2014bua}
are added with the help of~\FH, while further available fixed-order
results~\cite{Passehr:2017ufr,Borowka:2018anu} or resummation of large
logarithms~\cite{Hahn:2013ria,Bahl:2016brp,Bahl:2017aev} are
discarded; (3)~the assignment of the \cp-even light and heavy
doublets~$h^0,H^0$ and singlet~$s^0$ as well as the \cp-odd
doublet~$A^0$ and singlet~$a_s$ is determined from the largest
admixture of the corresponding tree-level state; (4)~the
mass~$m_{H^\pm}$ of the charged Higgs~$H^\pm$ substitutes~$A_\lambda$;
(5)~vacuum stability is determined from the tree-level Higgs potential
to a good accuracy, see~\citere{Hollik:201Xxxx}; (6) metastable vacua
are classified as long- or short-lived if the tunnelling time to the
deeper global minima is longer or shorter than the age of the
universe; (7)~constraints from charge and color-breaking
minima~\cite{Ellwanger:1999bv,Bobrowski:2014dla,Hollik:2015pra,Hollik:2016dcm}
can be taken over to the~\munmssm; (8)~we investigate tree-level
decays, but point out that loop-corrections can be
large~\cite{Nhung:2013lpa,Baglio:2015noa,Muhlleitner:2015dua,Domingo:2018uim};
(9)~the Higgs decays which depend on both~$\mu$ and~$\mue$
are~\mbox{$s^0\to h^0\,h^0$}, \mbox{$H^0\to s^0\,h^0$}, \mbox{$A^0\to
  s^0\,a_s$} and~\mbox{$s^0\to H^+\,H^-$}; (10)~in the neutralino,
chargino and sfermion sectors~$\mu$ and~$\mue$ only appear
as~\mbox{$(\mu+\mue)$} and~$\kappa\,\mue$ at the tree level, which is
why the shift~\mbox{$\mu+\mue\to\mue$} and
rescaling~\mbox{$\kappa\to\tilde{\kappa}=\kappa\,\mue/(\mu+\mue)$}
lead to identical spectra in the~\nmssm{} and~\munmssm{}.

\begin{figure}[t]
  \centering
  \includegraphics[width=.46\linewidth]{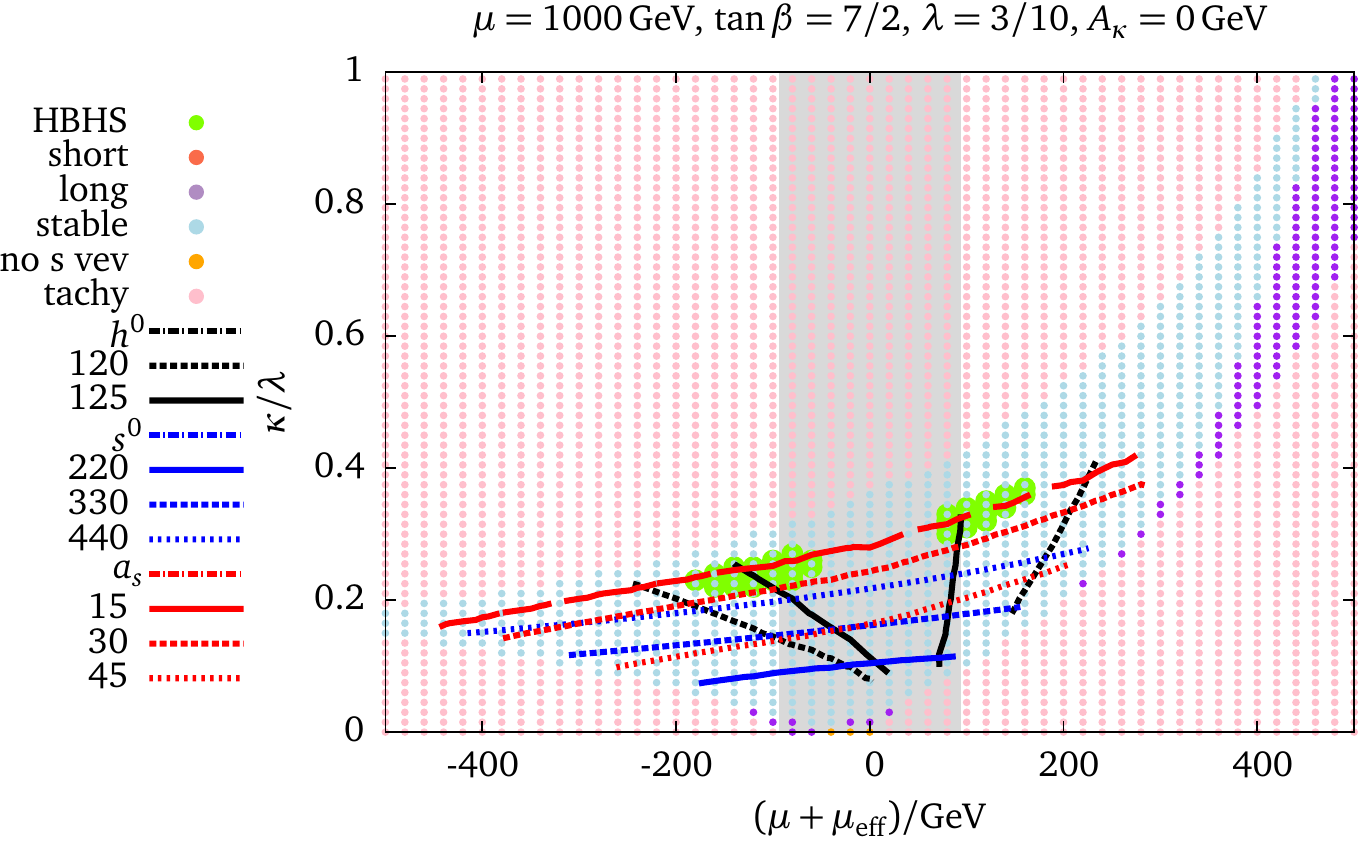}\hfill
  \includegraphics[width=.46\linewidth]{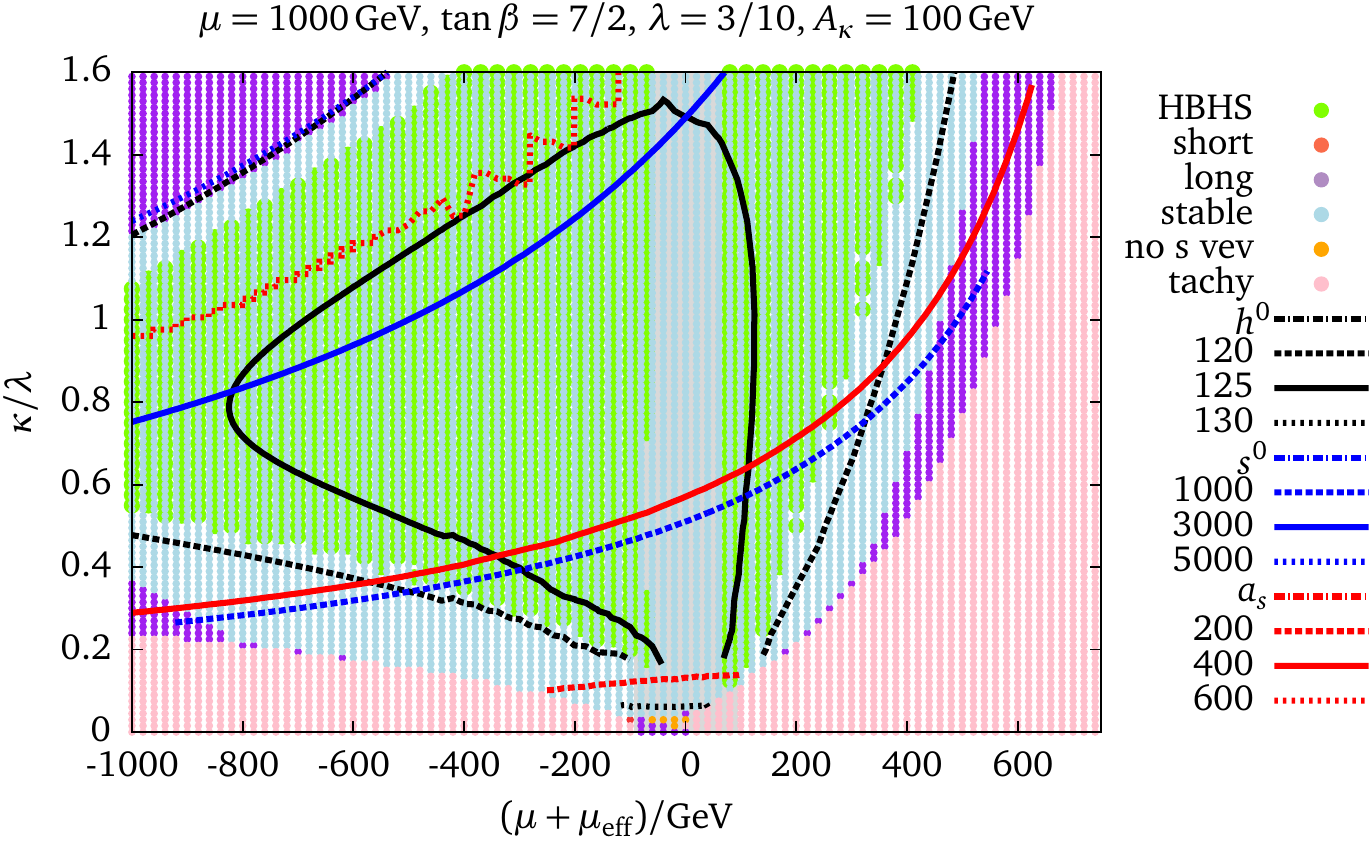}
  \caption{\label{fig:example}Mass contours of \sm-like Higgs (black),
    \cp-even singlet (blue) and \cp-odd singlet (red) in the
    plane~\(\kappa/\lambda\) versus~\((\mu + \mue)\), with~\(\lambda =
    3/10, \mu = 1000\,\GeV\) and varying~\(\kappa, \mue\) are
    shown. Left:~\(A_\kappa = 0\), right:~\(A_\kappa =
    100\,\GeV\). Furthermore,~\(\tan\beta = 7/2, m_{H^\pm}=800\,\GeV,
    m_t=173.2\,\GeV\). The soft-breaking SUSY parameters are the
    bilinear sfermion masses~$m_{\tilde{f}}=2\,\TeV$ and the trilinear
    sfermion mixings~$A_{f_3}=4\,\TeV$ for the third generation
    and~$A_{f_{1,2}}=0$ for the other generations. The state of the
    vacuum is indicated by the background colours: stable (light blue),
    metastable long-lived (purple), metastable short-lived (red),
    tachyonic (rose). No singlet vev exists in orange regions. The
    green region (HBHS) is allowed by \texttt{HiggsBounds} and
    \texttt{HiggsSignals}. The gray-shaded area is excluded by
    searches for charginos at LEP~\cite{Patrignani:2016xqp}.}
\end{figure}

\fig{fig:example} shows an example of our analysis in
\citere{Hollik:2018yek} for the Higgs spectrum and the vacuum
stability as a function of~$\kappa/\lambda$ and~$(\mu+\mue)$ with
fixed~$\lambda$ and~$\mu$. In this setup, variations in the
predicted~\sm-like Higgs mass are induced by admixture of the singlet
or loop corrections. The value of~$A_\kappa$ largely influences the
\cp-odd singlet mass, which explains the large tachyonic region in the
left plot. In the green area at the left side the singlet admixture to
the~\sm-like Higgs is reduced such that also the decay width
for~\mbox{$h^0\to a_s\,a_s$} becomes smaller. The endpoints of the
contours indicate a tachyonic Higgs at the tree or loop level. Each
boundary is parallel to the contours of the state which turns
tachyonic. The boundaries differ from the bounds of vacuum stability
due to the loop corrections.

\paragraph{Conclusions}

The extended parameter space of the~\munmssm{} compared to
the~\nmssm{} is investigated with respect to vacuum stability, Higgs
masses and decays, and neutralino phenomenology. Scenarios with the
potential to distinguish both models are proposed; differences
typically occur in the singlet sector. Some excluded scenarios in
the~\nmssm{} may still be viable in the~\munmssm.

\paragraph{Acknowledgments}

The authors are grateful to S.~Abel, P.~Basler, F.~Domingo,
K.~Schmidt-Hoberg, T.~Stefaniak and A.~Westphal for helpful
discussions.  The authors are supported by the DFG through a lump sum
fund of the SFB~676 ``Particles, Strings and the Early
Universe''. S.~P.\ acknowledges support by the~ANR grant
``HiggsAutomator''~(ANR-15-CE31-0002).

\newpage
\begingroup
\small
\setstretch{0}
\bibliographystyle{h-physrev}
\bibliography{NMSSM_inflation}
\endgroup

\end{document}

%% file: paperdef.tex
\usepackage[utf8]{inputenc}
\usepackage{epsf}
\usepackage{amsmath,amssymb}
\usepackage{graphicx}
\usepackage{listings}
\usepackage{array}
\usepackage{dcolumn}
\usepackage{mathtools}
\usepackage{scalefnt,ulem,soul}
\usepackage{cite}
\usepackage{color}
\usepackage[table]{xcolor}

\usepackage{etoolbox}

\makeatletter
\pretocmd\start@align
{%
  \let\everycr\CT@everycr
  \CT@start
}{}{}
\apptocmd{\endalign}{\CT@end}{}{}
\makeatother

\let\theparentequation\theequation
\patchcmd{\theparentequation}{equation}{parentequation}{}{}

\newcommand*{\nextParentEquation}[1][]{
  \refstepcounter{parentequation}
  \setcounter{equation}{0}
  \ifx\\#1\\\relax\else\parentlabel{#1}\fi
}

\makeatletter
\newlength{\fnhskip}
\renewcommand\@makefntext[1]{
  \settowidth{\fnhskip}{\@makefnmark}
  \leftskip=\fnhskip
  \hskip-\fnhskip
  \@makefnmark#1
}
\makeatother
\usepackage{setspace}

\usepackage[numbers,sort&compress]{natbib}
\makeatletter
\def\NAT@spacechar{\,}
\makeatother

\usepackage[
colorlinks=true
,urlcolor=blue
,anchorcolor=blue
,citecolor=blue
,filecolor=blue
,linkcolor=blue
,menucolor=blue
,pagecolor=blue
,linktoc=all
,unicode
,backref=false
]{hyperref}

\usepackage[all]{hypcap}

\usepackage[format=plain, margin=0.5cm, font=footnotesize, labelfont=bf]{caption}

\usepackage{cleveref}

\Crefname{figure}{Fig.}{Figs.}
\usepackage{placeins}

\let\theparentequation\theequation
\patchcmd{\theparentequation}{equation}{parentequation}{}{}

\apptocmd{\thebibliography}{\scriptsize}{}{}
\let\OLDthebibliography\thebibliography
\renewcommand\thebibliography[1]{
  \OLDthebibliography{#1}
  \raggedright
  \setlength{\parskip}{1pt}
  \setlength{\itemsep}{1pt plus 0.3ex}
}

\newcommand{\citere}[1]{Ref.\,\cite{#1}}
\newcommand{\citeres}[1]{Refs.\,\cite{#1}}

\newcommand{\code}[1]{\texttt{#1}}

\newcommand{\FH}{\code{FeynHiggs}}
\newcommand{\FA}{\code{FeynArts}}
\newcommand{\FC}{\code{FormCalc}}
\newcommand{\LT}{\code{LoopTools}}

\newcommand{\HBv}[1]{\code{HiggsBounds}--\code{#1}}
\newcommand{\HSv}[1]{\code{HiggsSignals}--\code{#1}}
\newcommand{\abbrev}{\scalefont{.9}}

\newcommand{\fig}[1]{Fig.\,\ref{#1}}

\newcommand{\sm}{{\abbrev SM}}

\newcommand{\mssm}{{\abbrev MSSM}}
\newcommand{\nmssm}{{\abbrev NMSSM}}
\newcommand{\munmssm}{{\abbrev $\mu$NMSSM}}

\newcommand{\cp}{{\abbrev $\mathit{CP}$}}

\newcommand{\drbar}{{\abbrev $\overline{\text{DR}}$}}

\newcommand{\GeV}{\textrm{GeV}}
\newcommand{\TeV}{\textrm{TeV}}

\newcommand{\mue}{\mu_{\text{eff}}}

\newcounter{notecount}
\makeatletter
\DeclareRobustCommand\em{%
  \@nomath\em \ifdim \fontdimen\@ne\font >\z@\scshape
  \else \slshape \fi}
\makeatother

\renewcommand{\emph}[1]{{\em #1}}